# Investigation on the critical dynamics of real magnetics models by computational physics methods

A.K. Murtazaev, V.A. Mutailamov

*Institute of Physics, Dagestan Scientific Center, Russian Academy of Sciences, Makhachkala, 367003 Russia*

*e-mail: vadim.mut@mail.ru*

The critical dynamics of classical 3D Heisenberg model and complex model of the real antiferromagnetic $Cr_2O_3$ is investigated with use of the method of molecular dynamics. The dynamic critical exponent $z$ are determined for these models on the basis of the theory dynamic finite-size scaling.

Despite of considerable successes reached to the present time in investigations of the critical dynamics of spin systems, the quantitative study of this problem still is one of actual problems of a modern statistical physics [1]. Some theoretical approaches available in this field, developed independent from each other and are based on completely different ideas. Among these approaches we shall mark the mode-coupling theory and theory of the dynamic scaling [1,2]. These theories have allowed qualitatively correctly explaining many experimental facts. It is possible to receive quantitative results for simple model systems only from the theoretical approaches with use of the renormalization group theory and ε-expansion theory [2,3].

Recently, the various approaches based on numerical methods are widely applied to study a dynamic critical behavior [4]. Note that using of numerical experiment methods allows investigating of complex real system models. The existing theoretical approaches do not allow it to make because of a strong dependence of dynamic critical phenomena from details of a Hamiltonian and type of potential. A method of the molecular dynamics together with the Monte-Carlo method is used in one of such approaches. So in papers [5,6] the approach was used for an investigation of the critical dynamics of some simple magnetic models.

The main point of this method consists in the following. The system is brought to a state of the thermodynamic equilibrium by the Monte-Carlo method. Then the system of differential equations of spins motion is solved. Using the dynamic finite-size scaling theory [7] and a procedure mentioned in [5,6] can be determine a value of the dynamic critical exponent $z$. As approbation we explored the Heisenberg ferromagnet model with the linear dimensions from $L=6$ up to $L=17$. For this model the Hamiltonian and the system of equations of spins motion with $k=x,y,z$ have the following view

$$H = -\frac{1}{2}\sum_{i,j} J_{ij}(s_i s_j) \quad, \quad \frac{\partial s_i^k}{\partial t} = \left[ J s_i^k \times \sum_j s_j^k \right], \quad t = t'(s/J\gamma), \quad |s_i|=1.$$

The solution of equations of spins motion allows to receive space-displaced, time-displaced correlation function

$$C^k(\vec{r}-\vec{r}',t) = \left\langle S_{\vec{r}}^k(t) S_{\vec{r}'}^k(0) \right\rangle.$$

Using space-time Fourier transform of correlation functions

$$S^k(\vec{q},\omega) = \sum_{r,r'} \exp[i\vec{q}(\vec{r}-\vec{r}')] \times \int_{-t_{cutoff}}^{+t_{cutoff}} \exp(i\omega t) C^k(\vec{r}-\vec{r}',t) \frac{dt}{\sqrt{2\pi}}$$

and define a characteristic frequency $\omega_m$ from the condition

$$\int\limits_{-\omega_m}^{+\omega_m} S^k(\vec{q},\omega)d\omega = \frac{1}{2}\int\limits_{-\infty}^{+\infty} S^k(\vec{q},\omega)d\omega,$$

it is possible to define dependence of a characteristic frequency $\omega_m$ versus the linear dimensions of system L. From the dynamic finite-size scaling theory follows, that

$$\omega_m(\vec{q},L) = L^{-z} f(qL).$$

So it is possible to define value of the dynamic critical exponent from the above mentioned equation maintaining a relation $qL=const$ for all systems.

The value obtained of the dynamic critical exponent $z$ is 2.49±0.09, that agrees good both with the theoretical value for three-dimensional isotropic ferromagnets ($z=5/2$) and results obtained in [6]. Fig.1 represents the dependence of a characteristic frequency versus the system sizes for a case $qL=2\pi$.

A study of the critical dynamics of the real magnetic materials represents the essential interest at the present stage of investigations of phase transitions and critical phenomena. This interest is conditioned by the absence of clear and unique view about the dynamic critical behavior of the real spin systems. In particular, there is a number of questions on a definition of the universality classes of the dynamic critical behavior.

Now we carry out investigations on the dynamic critical properties of models of complex real magnetic materials. In particular, the model of antiferromagnetic $Cr_2O_3$ are studied.

The Hamiltonian for $Cr_2O_3$ model have the following view

$$H = -\frac{1}{2}\sum_{i,j} J_1(\mu_i\mu_j) - \frac{1}{2}\sum_{k,l} J_2(\mu_k\mu_l) - D_0\sum_i (\mu_i^z)^2, |\mu| = 1,$$

where, according to the experimental data of [8], $J_1$ and $J_2$ are the parameters of the interaction of each spin with one nearest neighbor and three nearest neighbors, respectively ($J_2=0.45J_1$, $J_1<0$, $J_2<0$). The various relativistic interactions were fixed by the effective single-ion anisotropy $D_0>0$. We considered the following ratio between the anisitropy $D_0$ and exchange $J_1$

$$D_0/|J_1| = 2.5 \times 10^{-4},$$

corresponding to real $Cr_2O_3$ samples [9]. The temperature was the temperarure of phase transition: $T=T_c=0.466$. This value of $T_c$ we obtained from our previous investigations. Earlier we carefully investigated the static critical properties of $Cr_2O_3$ model with a calculation of statical critical exponents [10, 11].

The result for $Cr_2O_3$ model represent the value of z are $z=1.59\pm0.19$. We note that this value of dynamic critical index agrees with the theoretical estimates for isotropic ($z=d/2$, G model [2]) antiferromagnets.

Complete investigations of models of real ferro- and antiferromagnetic materials will allow to determine features of critical behaviour of models with weak additional interactions.

The investigation is supported by the Russian Foundation for Basic Research (projects №01-02-16103 and №02-02-06873-MAC) and by grant of a Commission of RAS for Supporting Young Scientists.

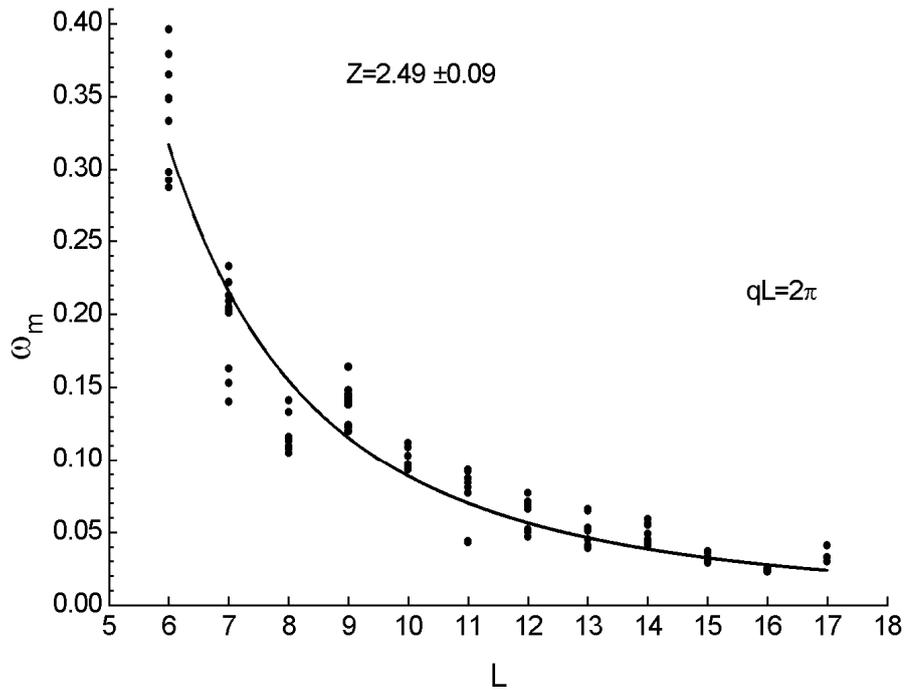

Fig.1. Dependence of a characteristic frequency versus the system sizes for a case $qL=2\pi$ for classical 3D Heisenberg ferromagnet model.